\documentclass[sigconf,balance=false]{acmart}

\usepackage{hyperref}
\usepackage{popets}
\usepackage{etoolbox}
\usepackage{tikz}
\usepackage{xcolor}
\usepackage{bm}
\usepackage{booktabs}
\usepackage{multirow}
\usepackage{caption}
\usepackage{subcaption}
\usepackage{tcolorbox}
\usepackage{colortbl}
\usepackage{soul}

\setcopyright{popets}
\copyrightyear{YYYY}

\acmYear{YYYY}
\acmVolume{YYYY}
\acmNumber{X}
\acmDOI{XXXXXXX.XXXXXXX}
\acmISBN{}
\acmConference{}
\colorlet{verylightgrey}{gray!15}
\NewDocumentEnvironment{takeaway}{ +b }{
    \setlength{\arrayrulewidth}{0pt}
    \setlength{\tabcolsep}{1mm}
    \vspace{0.9ex}\noindent\begin{tabular}{|>{\columncolor{verylightgrey}}c|}
    \hline%
    \parbox{\dimexpr(1\columnwidth -2\arrayrulewidth -2\tabcolsep)}{%
    \vspace*{0.9ex}#1\\[-1.7ex]
    } \\ \hline
    \end{tabular}
}

\newcounter{myfindingscounter}
\newcommand\takeawaytitle[1]{\textbf{Takeaway \refstepcounter{myfindingscounter}\themyfindingscounter\label{#1}:}}

\newcommand{\mypara}[1]{\noindent{\bf {\textsf{#1}}}}

\newcommand{\myparait}[1]{\noindent{\it {\textsf{#1}}}}

\newcommand*\emptycirc[1][0.65ex]{\tikz\draw (0,0) circle (#1);} 
\newcommand*\halfcirc[1][0.65ex]{%
  \begin{tikzpicture}
  \draw[fill] (0,0)-- (90:#1) arc (90:270:#1) -- cycle ;
  \draw (0,0) circle (#1);
  \end{tikzpicture}}
\newcommand*\fullcirc[1][0.65ex]{\tikz\fill (0,0) circle (#1);} 
\newcommand*\implemented{\tikz\fill[scale=0.4](0,.35) -- (.25,0) -- (1,.7) -- (.25,.15) -- cycle;} 

\author{Michael Wrana}
\affiliation{%
  \institution{University of Waterloo}
  \city{}
  \state{}
  \country{}
}
\email{mmwrana@uwaterloo.ca}

\author{Diogo Barradas}
\affiliation{%
  \institution{University of Waterloo}
  \city{}
  \state{}
  \country{}
}
\email{diogo.barradas@uwaterloo.ca}

\author{N Asokan}
\affiliation{%
  \institution{University of Waterloo}
  \city{}
  \state{}
  \country{}
}
\email{asokan@acm.org}

\begin{document}
\raggedbottom

\title[SoK: The Spectre of Surveillance and Censorship in Future Internet Architectures]{SoK: The Spectre of Surveillance and Censorship in \\Future Internet Architectures}

\begin{abstract}
Recent initiatives known as Future Internet Architectures (FIAs) seek to redesign the Internet to improve performance, scalability, and security. However, some governments perceive Internet access as a threat to their political standing and engage in widespread network surveillance and censorship.  In this paper, we provide an in-depth analysis of  the design principles of prominent FIAs in terms of their packet structure, addressing and naming schemes, and routing protocols to foster discussion on how these new systems interact with censorship and surveillance apparatuses.  Further, we assess the extent to which existing surveillance and censorship mechanisms can successfully target FIA users while discussing privacy enhancing technologies to counter these mechanisms.  We conclude by providing guidelines for future research into novel FIA-based privacy-enhancing technologies, and recommendations to guide the evaluation of these technologies.
\end{abstract}

\keywords{anonymous communications,  censorship, future Internet architectures, privacy, surveillance, traffic analysis}

\maketitle
\section{Introduction}
The dramatic growth of the Internet has enabled ubiquitous access to information, fostering seamless communication and promoting effective collaboration. Alongside it, multiple technological advances allowed network operators to more efficiently and effectively monitor and control the traffic that transits through their networks~\cite{DBLP:conf/ndss/Barradas00SRM21,DBLP:conf/sigcomm/SherryHSKRS12,DBLP:conf/ndss/DurumericMSBSBB17}. Unfortunately, these capabilities have empowered state-level actors to deploy large-scale surveillance and censorship mechanisms to monitor people's Internet activities or limit their ability to freely access and publish information~\cite{accessControlled,DBLP:conf/sp/NiakiCWHRCG20,DBLP:conf/ccs/RamanSKE20} (e.g., by blocking specific network addresses or websites).  

Although Internet surveillance and censorship mechanisms have substantially broadened and become more capable over the past decades~\cite{subramanian2011growth}, the deployment of effective privacy-enhancing tools remains a significant challenge. For instance, the traffic of popular anonymity networks such as Tor~\cite{DBLP:conf/uss/DingledineMS04} can be easily detected (and blocked) if not properly obfuscated~\cite{DBLP:conf/uss/DunnaOG18,DBLP:conf/ndss/MaticTC17}. Censorship evasion techniques like decoy routing~\cite{DBLP:conf/uss/WustrowWGH11} lack the support of major Internet Service providers (ISPs), and many alternatives based on covert channels demand esoteric software and networking know-how to be successfully used in practice~\cite{DBLP:conf/sp/TschantzAAP16}. Thus, preventing widespread surveillance and censorship remains an intricate challenge, in part due to the underlying architecture of the Internet (TCP/IP), and the specific methods used to send data between any two endpoints.

The original TCP/IP design principles focused on maximising usability and flexibility over security and privacy, due to the (perhaps optimistic) assumption that all involved parties would be trustworthy and well-behaved~\cite{DBLP:conf/fm/Vigna03}. In \S\ref{sec:bg-ip-censor} we show how, over time, the expectation of trust quickly disappeared, necessitating retroactive application of security enhancements as a means to protect against attacks ~\cite{DBLP:journals/concurrency/FenilK20, DBLP:journals/ton/HiltonHD22}. Furthermore, other Internet design principles that were originally considered carefully are now also being pushed to their limit (e.g., scalability issues caused by the growth of Internet-connected devices and the dearth of available IPv4 addresses).

To tackle the above challenges, multiple initiatives have pushed towards an overhaul of the Internet by applying the lessons learned throughout 30 years of practical network engineering experience. These initiatives, known as \textit{Future Internet Architectures} (FIAs), seek to remove legacy TCP/IP design constraints and improve performance, scalability, and mobility while adding much-needed security features. For instance, content-centric networks such as Named Data Networking (NDN)~\cite{DBLP:journals/ccr/0001ABJcCPWZ14} focus on making data contents named, addressable, and routable, while helping to provide integrity and authenticity at an architectural level (\S\ref{sec:ndn}). Other FIAs like eXpressive Internet Architecture (XIA) \cite{DBLP:journals/ccr/NaylorMAGKMBDHHKLOZLLSBAABDKKPPSSS14} focus their design principles on adaptability and future proofing the ever changing Internet.  Scalability, Control, and Isolation on Next-generation Networks (SCION)~\cite{SCION_book} overhauls TCP/IP routing by placing Autonomous Systems (ASes) into well-defined trust domains based on real-world legal and geopolitical boundaries, providing route control and failure isolation (\S\ref{sec:scion}).  

As of today, FIAs embody an attractive target for research, with ongoing efforts aimed at enhancing their practicality, e.g., by leveraging programmable networking hardware~\cite{DBLP:conf/acmicn/TakemasaKH21,DBLP:conf/conext/RuiterS21}, and experimenting with preliminary large-scale deployments~\cite{DBLP:conf/conext/KrahenbuhlTGKPH21,DBLP:conf/acmicn/UedaST22}. As these technologies mature, FIAs are gradually being deployed across various industries worldwide. For instance, the Swiss  financial and healthcare sectors are exploring the use of SCION to protect against cyberthreats, including denial-of-service or routing attacks~\cite{anapaya}. 

Despite these exciting leaps toward building novel, secure, and trustworthy internet architectures, existing work (including comprehensive surveys) has only examined FIA designs' ability to address broader security threats such as denial of service, cache attacks, or network intrusions~\cite{DBLP:journals/access/DingYD16, DBLP:journals/ccr/Fisher14, DBLP:journals/ieicet/Hasegawa13, mohammed2023future, DBLP:journals/cm/PanPJ11}, especially in content-centric networks~\cite{DBLP:journals/comsur/TouraniMMP18,DBLP:conf/lanman/GastiT18}, and NDN in particular~\cite{DBLP:journals/jcst/KumarSAS19, DBLP:journals/computers/HidouriHTHM22, DBLP:journals/corr/ChenM15f}. Some works analyzed security aspects associated with surveillance and censorship, but narrowed their focus on NDN~\cite{DBLP:journals/access/ShahLAB23, DBLP:journals/cm/ZhangYZNMLAZ18, DBLP:conf/milcom/ZhangWZ21, DBLP:journals/ccr/AbdelberiCKU13}, thus not considering looming threats to multiple disparate FIAs. For instance, while the content-centric nature of NDN may allow a state-level censor to block access to specific pieces of content cached in  routers~\cite{DBLP:journals/tifs/KimBVY17}, the highly-structured trust domains of SCION's inter-domain routing may also provide authorities with additional control over how network packets are allowed to traverse in, out, and through their networks~\cite{SCION_book}.

In this paper, we present a survey that aims to narrow the gap in the understanding of how novel FIA designs may impact the design and effectiveness of network surveillance and censorship mechanisms. In general, the extended functionality of intermediate routers and the additional packet header fields included in prominent FIA designs allow state-level actors to more easily enforce fine-grained surveillance and censorship policies (\S\ref{sec:censorship}). Furthermore,  the security enhancements proposed in FIAs do not prioritize defense against network surveillance and censorship. While subsequent research (\S\ref{sec:defense}) has introduced privacy-enhancing tools for countering surveillance and censorship in FIAs, many of such proposals have not seen a practical deployment and experimental evaluation, resulting in mere theoretical estimations of protection.  Finally, we suggest promising directions for continuing research into FIA-based privacy enhancing technologies, and offer specific guidelines for how best to evaluate these tools and techniques (\S\ref{sec:futurework}).
\section{Threat Model}
\label{sec:bg-ip-censor}
This section outlines a typical threat model in the Internet surveillance and censorship literature (which we also consider when performing our analysis of each FIA), and delivers a summarized outlook on prevalent approaches for implementing surveillance and censorship in the Internet Protocol (IP)~\cite{DBLP:journals/rfc/rfc791}. We refer to Appendix~\ref{bg:ip} for a description of the IP architecture. While this exposition suffices for grasping the primary vulnerabilities we identify on FIA-specific designs (\S\ref{sec:bg-fia}), we further deliver an exhaustive categorization of current surveillance and censorship techniques that focus on IP, and explore their potential applicability to FIAs (\S\ref{sec:censorship}).

\mypara{Network regions.} 
We consider two main network regions: (i) the \textit{censored} region, and (ii) the \textit{free} region. The censored region is assumed to be under the control of an omniscient adversary~\cite{DBLP:conf/sp/HoumansadrBS13} that can observe, store, interfere with, and analyze (e.g., resorting to sophisticated machine learning models) all the network traffic generated or received by any individual or organization located within the region. The free region consists of the part of the Internet that is not under the control of the adversary, or any other entity that aims to block Internet communications.

\mypara{Adversary capabilities.} The adversary we consider has direct control over the network infrastructure within its jurisdiction, including routers, switches, wires, DNS servers, etc. However, the adversary is unable to perfectly control the flow of packets across every link in their network. This exception is grounded on the fact that many real-world adversaries struggle to manipulate packet flows within their own networks~\cite{DBLP:journals/cacm/Goldberg14} and/or to enforce consistent country-wide traffic inspection and forwarding rules~\cite{DBLP:conf/www/WeinbergBC21}. As we will describe in \S\ref{sec:censorship}, some FIAs make it easier for network operators to have increased awareness and control over packet flows, increasing the risk of adversarial traffic manipulation. In addition, the adversary has no control over clients' endpoints (e.g., it cannot infect these with spyware~\cite{deibert2023autocrat}) and is computationally bounded, thus being unable to break cryptographic primitives used to encrypt network traffic.  The objective of the adversary is monitoring and filtering of Internet traffic, controlling what kinds of traffic should be allowed to flow within and across its network, and limiting which destinations network packets can travel to (be these either inside or outside the adversary's jurisdiction).
\section{Future Internet Architectures}
\label{sec:bg-fia}
This section presents an overview of the main design elements of six prominent FIAs and summarizes how these elements can either facilitate or hinder surveillance and censorship attempts conducted by state-level adversaries. While previous research~\cite{DBLP:journals/cm/PanPJ11,DBLP:journals/ieicet/Hasegawa13,DBLP:journals/ccr/Fisher14,DBLP:journals/access/DingYD16,mohammed2023future} mostly contrasted architectural differences and/or analyzed how the design of FIAs impact network performance, we place an emphasis on elements which are relevant from a privacy perspective, enriching our presentation with details about how these elements impact surveillance and censorship efforts.

As it will become clear in our analysis, and despite their multiple security improvements, FIA-specific designs and protocols leave open significant gaps on the privacy landscape that expose them to (or, in certain cases, escalate) surveillance and censorship threats.  While this section is centered on new potential vulnerabilities introduced by FIA-specific features, \S\ref{sec:censorship} analyzes whether FIAs can be subject to similar surveillance and censorship techniques (or variations thereof) that have been applied to IP in the past.

\begin{figure}
    \centering
    \includegraphics[width=0.95\linewidth]{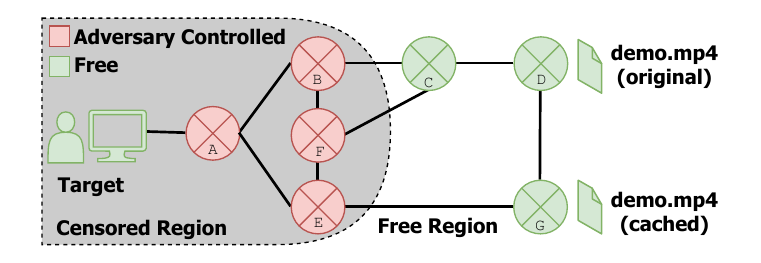}
    \vspace{-0.3cm}
    \caption{Example scenario: A user from router \texttt{A} attempts to access a video file from another host from router \texttt{D}.}
    \label{fig:demo-scenario}
    \Description[Example scenario: A user from router A attempts to access a video file from another host from router D.]{Example scenario: A user from router \textit{A} attempts to access a video file from another host from router \textit{D}.}
    \vspace{-0.3cm}
\end{figure}

\mypara{Scenario.} To motivate our discussion, we use the simplified network topology shown in Figure~\ref{fig:demo-scenario}. In our running example, a user located within a censored region wishes to watch a video that is hosted by the ``University of FIAs'' (uFIAs), a university located outside the censored region. We consider a network consisting of seven routers, labeled from \texttt{A} to \texttt{G}. The server hosting the video is located within the university's AS, which is distinct from the AS where the user is located. For the FIAs that allow routers to cache data (e.g., NDN), router \texttt{G} contains a cached copy of the video file.

\mypara{Choice of FIAs.} The set of FIAs we chose to focus on this paper is motivated by three main criteria: a) whether sufficient research and technical documentation to conduct a thorough analysis on architectural aspects exists; b) if any follow-up work has been conducted beyond the FIA's initial proposal, and; c) whether the FIA's design considerations can provide unique insights to our analysis. 
The majority of the FIAs we consider were prime candidates from the National Science Foundation (NSF) FIA awards~\cite{DBLP:journals/ccr/Fisher14}, which have led to the development of FIA prototypes beyond theoretical designs, and promoted a significant body of research on further analysis and refinements. We also include SCION~\cite{SCION_book}, an FIA that has  gained significant traction within the research community for the past few years, as well as NewIP~\cite{DBLP:conf/noms/ChenWLLJG20, DBLP:conf/hpsr/LiMD20, DBLP:conf/sigcomm/LiCCDM18}, a recent FIA that has stirred significant public discussion over network surveillance and censorship concerns. Below, we describe these architectures and their main design elements.

\subsection{Named Data Networking (NDN)}
\label{sec:ndn}
The Named Data Networking (NDN)~\cite{DBLP:journals/ccr/0001ABJcCPWZ14} project is an information-centric redesign of TCP/IP, which aims to improve address management, multicasting, and traffic regulation.

\mypara{Architecture.} In NDN, hosts specify data to collect from the network rather than hosts to be connected with. Thus, networks no longer aim to send a packet to a named destination but instead obtain named data from any location. Considering our running example (see Figure~\ref{fig:demo-scenario}), instead of requesting to connect with a specific host (e.g. uFIAs' website) via an IP address, requests are instead issued for a specific object (e.g. \texttt{uFIAs\_video.mp4}) via its unique name, and served from \textit{any} location (e.g. nearby router \texttt{C} which cached the requested video).

\mypara{Names and addresses.} Data names in NDN are the equivalent of addresses in IP, and are used by intermediate routers to forward packets to their destination.  A piece of data may be specified using a string similar to file or URL addressing (e.g. a video created at uFIAs might be named: \texttt{/uFIAs/videos/example.mp4}). 

\begin{figure}
\centering
\includegraphics[width=0.98\linewidth]{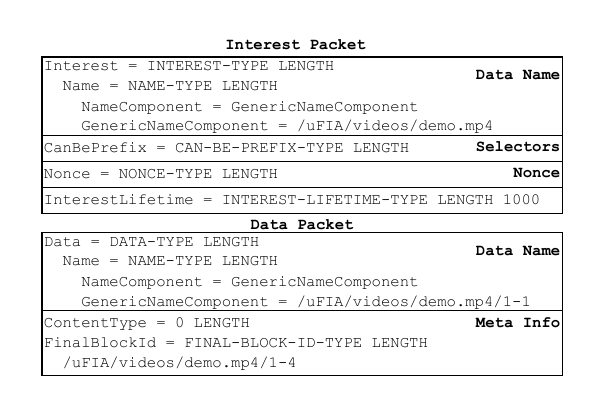}
  \vspace{-0.6cm}
  \captionof{figure}{Example NDN interest and matching data packet.}
  \label{fig:ndn-packet}
  \Description[Example NDN interest and matching data packet.]{Example NDN interest and matching data packet.}
  \vspace{-0.3cm}
\end{figure}

\mypara{Packet structure.} Sample packets based on the NDN specification are shown in Figure \ref{fig:ndn-packet}.  There are two types of packet in NDN: Interest and Data.  An interest packet is sent from a client to a server and represents a request for a given piece of named data (i.e., like HTTP GET). All interest packets are of variable length with two required fields, \texttt{name} and \texttt{nonce}, which combined can uniquely identify any packet.  Data packets contain a response to some specific interest packet and represent the requested resource.  

\mypara{Routing.}
In the control plane, NDN integrates well with traditional link-state or distance-vector routing protocols \cite{DBLP:conf/icccn/AfanasyevJYTXM017}.  Existing systems like OSPF and BGP can be adapted by treating names as a series of components and performing longest prefix matching using some separator (e.g. \textquoteleft /\textquoteright).  The data plane of NDN differs greatly from that of TCP/IP with each router containing three tables of information (versus just the FIB in TCP/IP).  The \textit{pending interest table} contains a list of all the interests that this router has forwarded, but not received a data packet for.  Each entry contains the name of the data, along with incoming and outgoing interfaces.  The \textit{forwarding information base} is a routing table that maps names to interfaces.  The \textit{content store} is a temporary cache of data packets that have been sent through this router. When a router receives a data packet, it determines which interfaces it should be sent to, based on the associated pending interest table entry. Each router independently caches data packets within its content store (CS).  If an interest packet arrives and the name matches a content store entry, the router can immediately send a cached copy to the client.

\mypara{Potential vulnerabilities.} NDN exhibits three characteristics that may make it more vulnerable to surveillance and censorship efforts.  (1) An adversary can \textit{analyze the data names} being requested in NDN packet headers and filter those packets based on some matching criteria. In this sense, data name filtering is similar to DNS or IP blocking. (2) One of the built-in protections against data integrity attacks in NDN is the requirement that all data must be signed by its original creator.  \textit{Data signatures} open up censorship opportunities for state-level adversaries seeking to \textit{block a particular information source}. Censors can create a key blocklist (e.g., block all data signed with uFIA's public key).  By using a key blocklist, state-level adversaries are capable of preventing access to content without relying on the name of a given piece of data. Further, key blocklists can also be used to target a specific client within the censored region and prevent them from sending information to the free network region.  (3) The \textit{data cache} present in content-centric networks like NDN can be exploited by state-level censors using, for instance, \textit{cache-enumeration} and \textit{timing} attacks. These attacks would allow censors to monitor and/or filter the pieces of data accessed by clients located within the censored region. 
\subsection{MobilityFirst (MF)}
\label{sec:mf}
MobilityFirst (MF) \cite{DBLP:conf/aintec/SeskarNNR11} redesigns the network-layer using a content centric architecture and is designed to support the shift from static hosts (e.g., desktop computers using WiFi) to mobile objects (e.g., smartphones using cellular data).

\mypara{Architecture.} Similarly to NDN, MobilityFirst follows a content-centric architecture aimed at obtaining data rather than connecting hosts \cite{DBLP:journals/sigmobile/RaychaudhuriNV12}.  All objects (i.e., devices, services, files, etc.) have both a globally unique object identifier and a network address.  Unique object identifiers are long-lasting public keys assigned to network objects by name certification services.  Network addresses are short-lasting routable addresses used to send packets to their destination.  

\mypara{Names and addresses.} In MobilityFirst, human-readable names, globally unique object identifiers, and network addresses are all stored and managed independently.  Globally unique object identifiers are assigned by a decentralized name certification service and represent specific devices (similar to a MAC address).  Objects are assigned identifiers using name certification services for their respective type (i.e. separate certification services for content and devices). Each piece of content's unique identifier is bound to a network address during routing using a \textit{global name resolution service}.  Network addresses are flexible and represent an object's current connection within the network. For example, a phone has one globally unique ID but may have two network addresses for its cellular and WiFi connections.  If the WiFi connection is lost, packets can be automatically re-directed to the cellular network address using the phone's unique ID.  Network address resolution in MobilityFirst is implemented as a distributed hash table \cite{DBLP:journals/ton/StoicaMLKKDB03, DBLP:conf/sigcomm/RatnasamyFHKS01}.

\begin{figure}
\centering
\includegraphics[width=\linewidth]{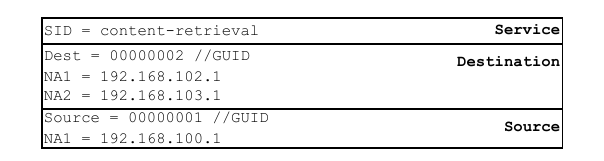}
  \vspace{-0.7cm}
  \captionof{figure}{Example MobilityFirst packet header.}
  \label{fig:mf-packet}
  \Description[Example MobilityFirst packet header.]{Example MobilityFirst packet header.}
\vspace{-0.2cm}
\end{figure}

\mypara{Packet structure.} Figure \ref{fig:mf-packet} depicts a sample MobilityFirst packet header.  MobilityFirst packets have three main components.  The source and destination addresses are represented as both network addresses and globally unique object identifiers.  Network addresses can be changed mid-path automatically while globally unique IDs are fixed.  Multiple network addresses can be assigned to a single globally unique ID (e.g., a phone connecting to the internet via cellular data and WiFi).  The service ID field specifies how the packet should be delivered (e.g., unicast, multicast, anycast). 

\mypara{Routing.} Instead of traditional link-state and distance-vector routing, MobilityFirst uses a storage-aware routing protocol based on a cache-and-forward architecture~\cite{paul2008cache, DBLP:conf/lanman/GopinathJMR10} which combines link-state routing with a delay-tolerant network architecture~\cite{DBLP:journals/jsac/FallF08}.  Further, due to the nature of names and addresses in MobilityFirst, inter-domain routing can no longer be performed using BGP.  As an alternative, Mukherjee \textit{et al.} \cite{DBLP:journals/cn/MukherjeeSVR17} propose edge-aware inter-domain routing which replaces individual routers and ASes with aggregated nodes and virtual links. MobilityFirst allows routers to cache packets to improve efficiency and stability when faced with network disconnections or bandwidth constraints. 

\mypara{Potential vulnerabilities.} It is possible that censors might exploit MobilityFirst's global name resolution service (more precisely, the dynamic network address resolution protocol) to \textit{change the final destination of a packet} while it is in transit. Crucially, the design choice of supporting in-flight address changes prevents IPSec-like defenses from guaranteeing address integrity.  Specifically, a state-level adversary with control over the network routers within the censored region could manipulate the resolution process to re-direct packets to a location of their choosing. Such an attack is akin to an adversary manipulating the IP-link address resolution in TCP/IP.  This could force packets through a censorship device which, returns forged replies or no response at all. 

Furthermore, since MobilityFirst is also a content-centric network (like NDN), they share some specific vulnerabilities on what it concerns the usage of in-network caching. For instance, similarly to NDN, censors are able to \textit{add specific data names to a blocklist} or scan network caches for collecting \textit{evidence that clients requested a given content}. This tracking can be made even easier as  MobilityFirst's network caches also store information about the original network interface from where a data request was received.
\subsection{NEBULA}
\label{sec:nebula}
The goal of NEBULA \cite{DBLP:journals/ccr/AndersonBBCCCFHIKLLMNSSRWWY14} is to support ongoing developments in cloud computing by providing a secure network infrastructure~\cite{DBLP:conf/fia/AndersonBBCCCFHIKLLMNSSRWWY13}.

\mypara{Architecture.} 
NEBULA re-designs three main components of the Internet: routing policy, data plane, and control plane.  One of NEBULA's innovations is the use of declarative networking \cite{DBLP:journals/cacm/LooCGGHMRRS09} to pass and enforce policies from applications in the network layer.  To improve internet routing, the NEBULA authors propose a set of ultra-reliable high-performance routers \cite{DBLP:journals/internet/AgapiBBCKMPSR11} designed to directly connect data centers with each other. NEBULA re-designs the data plane with a focus on resilience, mutual agreement to participate, and policy enforcement.  NEBULA introduces a new control plane with virtual and extensible networking techniques designed to provide policy specification, path construction, and address assignment.

\mypara{Names and addresses.}
NEBULA introduces an additional service abstraction layer \cite{DBLP:conf/nsdi/NordstromSGKAKRF12} between the network and transport layers that enables applications to communicate directly using service names. The new layer divides addresses into three components: the service name, the address, and the flow.  A service name is a group of processes that all offer the same service.  Addresses identify a host interface, and flows indicate the specific flow associated with a socket.  The service abstraction layer maps service addresses in packets to network addresses with a service table. Once a destination address has been resolved, the client checks its cache for a saved path.  If no path is saved, the client can query consent servers until either a path is found or an error is returned.

\begin{figure}
\centering
\includegraphics[width=0.98\linewidth]{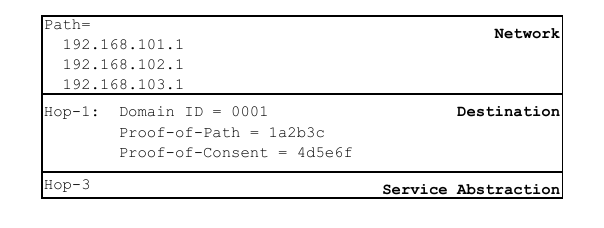}
  \vspace{-0.6cm}
  \captionof{figure}{Example NEBULA packet header with ICING.}
  \label{fig:nebula-packet}
  \Description[Example NEBULA packet header with ICING.]{Example NEBULA packet header with ICING.}
  \vspace{-0.3cm}
\end{figure}

\mypara{Packet structure.} A sample NEBULA packet is shown in Figure~\ref{fig:nebula-packet}.  NEBULA expands the network layer with support for ICING \cite{DBLP:conf/conext/NaousWNMMS11} packets that provide path verification.  In the ICING protocol, each packet contains three pieces of information for every router on the chosen path: A node ID and its corresponding tag, a proof of consent, and a proof of provenance.  The Node IDs are listed sequentially followed by a constant-length verifier that aggregates all the proof tokens.  The proof of consent is a cryptographic token that proves the router's provider agrees to the selected path. The proof of provenance verifies that packets traveled through the pre-arranged domain in the correct order.  When a packet arrives at an intermediate router, it cryptographically verifies the embedded tokens before forwarding the packet to its destination. NEBULA also appends a service abstraction layer header between the new network layer header and the transport layer header.  

\mypara{Routing.} NEBULA virtual and extensible networking techniques provide a control plane architecture using declarative networking \cite{DBLP:journals/cacm/LooCGGHMRRS09}.  These techniques also define an API called Serval \cite{DBLP:conf/nsdi/NordstromSGKAKRF12} for specific policy-based service requests.  In the NEBULA control plane, administrators provide high-level specification for route policies while avoiding specific implementation details.  NEBULA provides two tools to assist with routing: (1) BGP-style global reachability (2) an interface to the data plane for specific policy-based path generation.  In turn, the new data plane architecture of NEBULA uses a path verification mechanism \cite{DBLP:conf/conext/NaousWNMMS11} in which the action of forwarding a packet to its next-hop router is distinct from routing (where topology discovery and path selection occur).

\mypara{Potential vulnerabilities.}  Whereas NEBULA provides guaranteed route integrity and proof to both end-users that packets followed a pre-approved path, clients located within a censored region must cross routers controlled by the adversary. Thus, adversaries can modify router behaviour by taking advantage of proof-of-consent tokens. A state-level adversary can simply block or filter packets that are not approved by their own service (e.g., block all packets that contain proof tokens from routers serving Google servers).  Compared with other FIA designs, NEBULA is fairly resilient towards in-network attacks and tampering by state-level adversaries.  However, these built-in defenses can be perceived as a double-edged sword because it complicates the deployment of anonymous communication and censorship evasion tools~\cite{DBLP:journals/comsur/AmbrosinCCGT18}.
\subsection{eXpressive Internet Architecture (XIA)}
\label{sec:xia}
eXpressive Internet Architecture (XIA)~\cite{DBLP:journals/ccr/NaylorMAGKMBDHHKLOZLLSBAABDKKPPSSS14, DBLP:conf/hotnets/AnandDHLLMWAABSS11} aims to allow for network layer changes to be deployed without the need to replace existing protocols (e.g., XIA can co-exist with IP).

\mypara{Architecture.} XIA leverages three main concepts: principals, fallback addressing, and intrinsically secure identifiers.  A principal in XIA is an abstraction of hosts in the current internet.  Principals are identifiable senders or receivers of a packet, and can represent a host, service, or piece of content. Fallback addresses support integrating new network-layer protocols. If some legacy router encounters an unknown principal, fallback routes describe alternative actions for the router to take.  XIA's Self-certifying identifiers are used for all principals to bootstrap trust management. Security properties can be specific to each principal, allowing it to verify that communications are established with the intended target, avoiding any need for external tools.

\mypara{Names and addresses.} 
XIA provides a set of built-in principals but supports an arbitrary number of future options.  There are four basic principal identifiers in XIA: host IDs define who is on either end of a communication, service IDs define what an entity does (e.g., serving webpages), content IDs define what a piece of content is, and network IDs verify that a principal is communicating with the correct network. XIA addresses are represented as directed acyclic graphs (DAGs) of the path between source and destination principals.  Every address must have a source and destination represented as some principal identifier.  If the primary route is unreachable due to network changes or router incompatibility, fallbacks are included as additional edges in the graph.  Breaking the DAG into sub-graphs supports longest-prefix matching.

\begin{figure}
    \centering
    \includegraphics[width=0.95\linewidth]{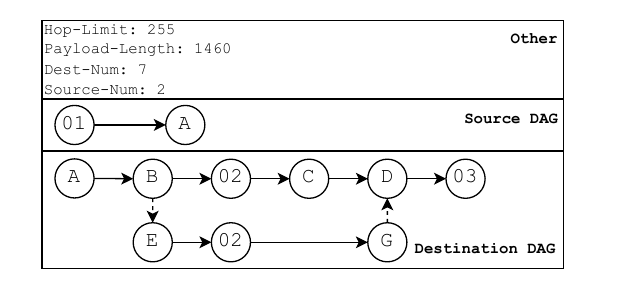}
    \vspace{-0.5cm}
    \caption{Example XIA packet header.}
    \label{fig:xia-packet}
    \Description[Example XIA packet header.]{Example XIA packet header.}
    \vspace{-0.3cm}
\end{figure}

\mypara{Packet structure.} An example of an XIA address and its associated DAG is shown in Figure~\ref{fig:xia-packet}.  XIA packet headers must contain the source and destination addresses along with an optional accountability address (omitted in Figure \ref{fig:xia-packet}) used to verify whether the packet is following the correct approved path.  Additionally, XIA packet headers contain the total number of source and destination nodes, time-to-live, and payload length fields.  

\mypara{Routing.} 
Routing in XIA operates similarly to IP.  The network ID nodes close to the source are used as a prefix for longest-prefix matching algorithms when determining the correct outgoing port.  The host ID indicates a specific device, and the content ID represents a socket on that host (similarly to a TCP port).  Distant routers forward packets to a network ID then local routers send packets to a specific host.  Since the entire destination is not required by distant routers, the specific service or content ID can be encrypted until a packet has arrived at the destination network.  XIA is flexible enough to support any inter-AS routing protocol, although the authors suggest SCION as an effective solution (\S\ref{sec:scion}).

\mypara{Potential vulnerabilities.} 
XIA packets include the complete path that a packet follows from a principal's source interface to the destination network and the specific process at the destination server.  Censors could use this information to determine details about an encrypted packet or connection.  Interestingly, XIA's design includes a countermeasure where part of the destination address could be encrypted so it can only be read by the destination network. Suppose a user attempts to covertly communicate with a censored target, the \textit{University of FIAs (UFIAs)} by encrypting their packet path (shown as the Destination DAG in Figure \ref{fig:xia-packet}) beyond the \texttt{N02} node.  Unfortunately, an adversary can \textit{deduce the encrypted portion of a packet's path}.  To launch their attack, the adversary first requests similar files from uFIAs, and records the fallback addresses and path head (i.e., the unencrypted portion of the complete path) between \texttt{A} and \texttt{N02}.  Then, the adversary enacts an aggressive filtering policy where all users attempting to send packets towards \texttt{N02} are blocked, under the assumption that they are attempting to communicate with uFIAs.  The filtering policy blocks the aforementioned user who is attempting to evade censorship by path encryption.  The efficacy of blocking partially encrypted paths might be limited by two factors.  First, if paths towards uFIAs are not unique, the censor cannot identify the intended destination from just the path head.  Second, if there are multiple destinations within \texttt{N02} that have the same path head and \textit{should not} be censored, the policy would be too aggressive and thus ineffective.

An adversary may also be able to \textit{exploit the fallback addressing} feature to \textit{route packets through specific locations} and prevent access to selected services.  Since network routers are given complete control over whether to follow the intended path or some provided fallback address, an adversary may configure its routers to prevent packets from following the network path that the source intended them to follow. Further, the adversary may insert their own fallback address into packet headers, providing the adversary with path manipulation capabilities that would provide a tight control over the traffic flowing within its jurisdiction.
\subsection{SCION}
\label{sec:scion}
SCION \cite{DBLP:journals/corr/BarreraRSP15, DBLP:conf/sp/ZhangHHCPA11, SCION_book} stands for Scalability, Control, and Isolation on Next-generation Networks.  The goal of SCION is to redesign the inter-AS routing protocols of the current internet.

\begin{figure}
    \centering
    \includegraphics[width=\linewidth]{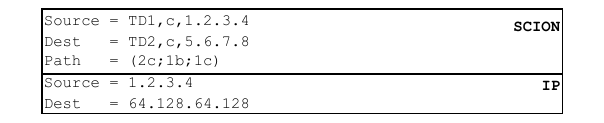}
    \vspace{-0.8cm}
    \caption{Example SCION packet header.}
    \label{fig:scion-packet}
    \Description[Example SCION packet header.]{Example SCION packet header.}
    \vspace{-0.3cm}
\end{figure}

\mypara{Architecture.} 
An Autonomous System (AS) is a large network of internet devices managed by a single internal routing policy.  SCION organizes the many ASes of the internet into trust domains which contain ASes that agree on mutual trust based on some shared framework (e.g., geopolitical, legal, etc).  In SCION, there should be a limited number (\textless 500) of top-level trust domains indexed and identified with a human-readable name.  Each trust domain is administered by a core which contains ASes that perform additional roles including bootstrapping and routing policy management.  In SCION, ASes are organized hierarchically within their trust domain based on customer-provider relationships.  Each AS may belong to multiple trust domains, and is allowed to peer with other ASes freely outside the domain.

\mypara{Names and addresses.} SCION addresses are represented as a combination of trust domain, AS, and host addresses.  Each trust domain address should be globally unique, while AS addresses are unique within each trust domain, and host addresses must be unique within the specified AS.  Host addresses are variable-length and can be IPv4 or IPv6, with an additional service ID used for control plane requests as part of the control message protocol (akin to ICMP).

\mypara{Packet structure.} A simplified sample packet for SCION is shown in Figure \ref{fig:scion-packet}. SCION packets are generated by appending an additional header to traditional IP packets.  Every SCION packet has a mandatory common header containing (among other information) the packet length, source/destination address type, and current path position.  Remaining fields include the source and destination address, as well as detailed path information. 

\mypara{Routing.} SCION divides inter-AS routing into two categories: inter-trust-domain and intra-trust-domain, while leaving existing intra-AS routing untouched.  
Routes in SCION are divided into up-paths, down-paths, and core-paths. Up-paths and down-paths are routes to and from an AS and its trust domain core.  Core-paths are established between top-level trust domain core ASes, and can be obtained with a deterministic link-state protocol as the number of such core ASes is limited.

Up-paths and down-paths are obtained through the use of \textit{path construction beacons}, which represent a path segment (similar to BGP's \texttt{as\_path} attribute) and are used as part of the process to construct a complete end-to-end path between two hosts.  Beacons are periodically created by core ASes and propagated downwards through the trust domain following customer/peer relationships.  Upon receiving a beacon, each AS adds itself along with its ingress and egress interfaces to the existing path.  Any peering links between two ASes are also specifically indicated within the beacon.  Each AS maintains a list of paths obtained from beacons and chooses a subset of up-paths and down-paths it prefers to use.  The up-paths are stored within each AS independently, while down-paths are sent to the trust domain core ASes.

\mypara{Potential vulnerabilities.}  Adversaries may take advantage of different SCION sub-systems to help obtaining granular route control throughout their network, facilitating the enforcement of surveillance and censorship mechanisms. First, adversaries that have control over multiple ASes can \textit{manipulate the SCION path construction process} via path interposition attacks~\cite{SCION_book}. In these attacks, the adversary can block traffic between two ASes they control, and force all traffic to be re-routed through another AS. An adversary may launch these attacks to force traffic to transit through routers designed to enforce censorship policies. Through a combination of these two techniques, the adversary can establish SCION-enforced cryptographically verifiable \textit{granular route control} throughout their entire network.  Suppose a user attempts to evade this policy using path construction beacon theft \cite{SCION_book}.  In this attack, the user injects an AS into their packet's path which contains a peering link into the free region to evade censorship efforts at a particular router.  Unfortunately, SCION's ``next-hop'' interface in the path construction beacon would prevent this user's path from being valid, thus enforcing the state censorship policy.  The only way to evade path-based censorship in SCION is for the target to establish an out-of-band communication channel to exchange valid path information.

A third SCION design element that could be exploited by an adversary is the tight hierarchical organization that arises from SCION's AS model. This stands in contrast to well-known censorship systems such as the Great Firewall of China, where different ISPs receive censorship guidelines, leading to inconsistent execution. As a result, certain methods for evading censorship might be effective in one location but not in another~\cite{DBLP:conf/www/WeinbergBC21}.  The trust domain architecture of SCION offers state-level adversaries the ability to \textit{force traffic through a single location} and greater control over ASes. 
\begin{figure}
\centering
\includegraphics[width=0.98\linewidth]{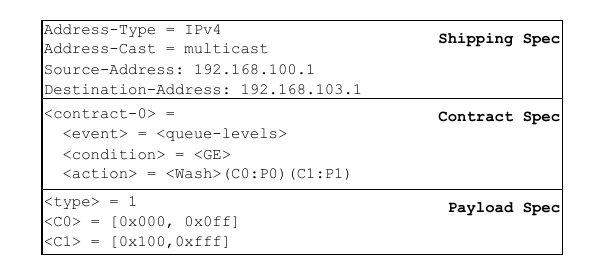}
  \vspace{-0.5cm}
  \captionof{figure}{Example NewIP packet header.}
  \label{fig:newip-packet}
  \Description[Example NewIP packet header.]{Example NewIP packet header.}
  \vspace{-0.3cm}
\end{figure}

\subsection{NewIP}
\label{sec:newip}

NewIP~\cite{DBLP:conf/noms/ChenWLLJG20, DBLP:conf/hpsr/LiMD20, DBLP:conf/sigcomm/LiCCDM18} is based on the concept of Big Packet Protocol \cite{DBLP:conf/sigcomm/LiCCDM18}, where new information is appended into packets that contains metadata and commands that provide additional guidance.

\mypara{Architecture.}  NewIP modifies and expands the network layer with its own protocol, providing three main features. First, flexible IP addressing and length with support for IPv4, IPv6, and any future addressing standards.  Second, a semantic definition of the IP address to identify physical and virtual objects.  Third, a user definable packet header allowing hosts to specify custom functions to be applied to their outgoing packets by other network devices.

\mypara{Names and addresses.} 
Different types of network-layer addresses such as IPv4 or IPv6 can be used as part of the NewIP architecture.  The NewIP header allows for seamless integration of different network structures into NewIP, and backwards compatibility with existing addressing schemes.  Furthermore, alternate addressing formats like LiRA \cite{stoica1998lira} (which supports a differential service model using resource tokens) could also be seamlessly integrated.

\mypara{Packet structure.} 
The New IP packet header (Figure \ref{fig:newip-packet}) consists of three specification sections: shipping, contract, and payload.  The shipping specification is a required portion of the header that contains an address of flexible length and type.  The Shipping-Spec header contains address-type (e.g., IPv4 = 01), address-cast (e.g., multicast), source address, and destination address sections.  The contract specification describes a formal service specification that may include network capability, actions, and accounting information.  The payload specification allows a NewIP packet to carry contextual information about the packet's content. 

\mypara{Routing.} NewIP's additional routing capabilities are primarily introduced via the contract specification header, which is composed of an arbitrary number of clauses that each represent an event, condition, or action. Events represent local occurrences within the network which may affect a packet's behavior.  A condition represents a logical operator to be performed on the event, and each available action is chosen from a pre-determined set that is known and shared by all nodes in the network.  Packets in NewIP can be \textit{traditional} (i.e., no payload specification) or \textit{qualitative}, in which the payload specification describes quality, semantic, or other information.  This can be used to further enhance communications according to the NewIP network objectives.

\begin{table}[t!]
\caption{Summary of FIAs' features most relevant to censorship and surveillance.}
\label{tab:arch-table}
\vspace{-0.3cm}
\resizebox{\linewidth}{!}{
\begin{tabular}{llllllll}
\hline
\textbf{Type} & \textbf{NDN} & \textbf{SCION} & \textbf{MF} & \textbf{XIA} & \textbf{NewIP} & \textbf{NEBULA} & \textbf{TCP/IP} \\ \hline
\multicolumn{8}{c}{\textbf{Architecture Features}} \\
\textbf{Design} & Content & Host & Content & Hybrid & Host & Host & Host \\
\textbf{Caching} & Routers & None & Routers & Routers & None & None & None \\
\textbf{Resolution} & NDNS & DNS & GNRS & DNS & DNS & DNS & DNS \\ \hline
\multicolumn{8}{c}{\textbf{Packet Format Features}} \\
\textbf{Routes} & None & Partial & None & Yes & None & Yes & None \\
\textbf{Addresses} & Destination & Both & Both & Both & Both & Both & Both \\
\textbf{Content} & Name & None & None & Name & None & Name & None \\
\textbf{Service} & None & None & Yes & Yes & Partial & None & None \\ \hline
\multicolumn{8}{c}{\textbf{Routing Features}} \\
\textbf{Data Plane} & FIB/PIT/CS & FIB & GUIDs & XIDs & FIB & FIB & FIB \\
\textbf{Inter-AS} & BGP & SCION & EAID & SCION & BGP & NVENT & BGP \\
\textbf{Intra-AS} & OSPF & OSFP & GSTAR & OSPF & OSPF+ & NVENT & OSPF \\
\textbf{Multi-Path} & Yes & Yes & Yes & Yes & Yes & Yes & Yes \\ \hline
\end{tabular}
}
\end{table}

\mypara{Potential vulnerabilities.} NewIP was proposed by Huawei with support from China Mobile, China Unicom, and CAICT~\cite{ISOC-huaweiNewIP}, sparking discussion about individual rights to privacy and free access to information~\cite{FinancialTimesNewIP}. NewIP also earned the support of other countries known to filter Internet access, e.g., Russia~\cite{FinancialTimesITUProposal}. Below, we discuss some  NewIP's characteristics that could potentially allow network operators to more easily conduct surveillance and censorship.

Adversaries may attempt to \textit{modify the contracts} within NewIP packet headers to simplify surveillance and censorship efforts. In part, this is due to the fact that NewIP does not include any provision for the protection of packet headers, which must be transmitted in plaintext to allow for routers to use the information contained therein.  The censor can either add contracts directly to the packet header, meaning even external routers in the wider internet would provide them with tracking data.  Alternatively, they could trace and track packets within their region regardless of what (if any) contract is listed in the header. Many contract-based attacks (or combinations thereof) are possible.

More precisely, the adversary can add a \textit{PktTrace} action to a flow that is suspected of containing prohibited data. Routers outside the censor's jurisdiction would pass information about the path followed and time spent at each node back to the censor. The trace information could be directly used to block the flow (if it traveled to a prohibited host) or provide further data for classification. Depending on the implementation of \textit{PktTrace}, this may also provide flow metadata to the censor (e.g. size, timing, etc).  Alternatively, latency bounds could be added with \textit{BoundedLatency(t)} actions such that they are dropped after a pre-specified time.  Any communication using a decoy routing or onion routing approach that includes non-negligible latency would be consistently dropped.

\begin{figure}
\centering
\includegraphics[width=\linewidth]{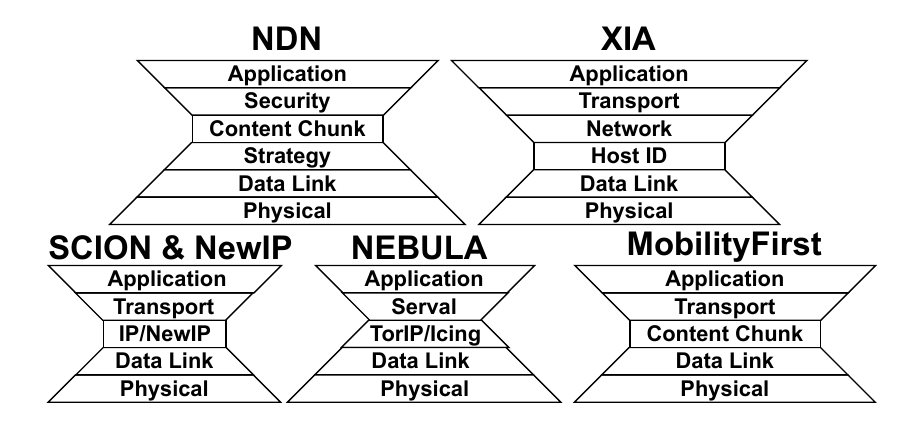}
  \vspace{-0.8cm}
  \captionof{figure}{Comparison of the hourglass models of each FIA.}
  \label{fig:hourglass}
  \Description[Comparison of the hourglass models of each FIA.]{Comparison of the hourglass models of each FIA.}
  \vspace{-0.3cm}
\end{figure}

\subsection{Summary}

Table \ref{tab:arch-table} showcases the high-level design choices of each FIA considered in our analysis (comparing them with TCP/IP) and how these choices affect the way data flows through the network. In Fig. \ref{fig:hourglass} we show an overview of how each FIA modifies the OSI hourglass model.  We now summarize different aspects tied to the architecture, packet format, and routing decisions that shape each FIAs design.

\mypara{Architecture features.} 
The \textit{Design} aspect distinguishes between host-centric and content-centric networks.  We discussed three FIAs that support content-centric designs, and three that maintain TCP IP's host model. Architectures with \textit{Caching} are built around the ability to store data within network routers. Only the content-centric designs allow native support this concept. Some FIAs introduce new \textit{Name Resolution} services, while others rely on the existing DNS infrastructure. MobilityFirst is the only FIA that introduces a completely new way of resolving hostnames, while the other architectures integrate or rely on DNS in a significant way.

\mypara{Packet format features.}
Some FIAs' packet structures provide \textit{Routes}, where a packet contains a list of all the routers it will go across. SCION's and XIA's packets have a loose structure that provides a list of routers or networks that a packet will traverse through, while NEBULA uses a cryptographically enforced policy where packets must follow a pre-defined route embedded in the header. Almost all FIAs list both the source and destination \textit{Addresses} in their header.  Instead, NDN only requires the destination (or data name) to be listed in a packet. \textit{Content} represents whether the packet headers contain any information about their payload.  Here, content-centric networks represent what their payload contains in the destination address as a data name or ID. Some packet structures also include information about the \textit{Service Type} associated with the flow (e.g., MobilityFirst, XIA). FIAs that integrate the service and network layers (e.g., NewIP) require packet headers to record the exact way a packet travels through the network.

\mypara{Routing features.}
The \textit{Data Plane} concerns the primary systems used to facilitate  packet forwarding. All the content-centric network designs introduce unique new ways to forward packets. \textit{Inter and Intra-AS} routing indicates either the proprietary technique for routing introduced with the FIA or the algorithm suggested by the authors.  Here, NEBULA, SCION, and MobilityFirst propose their own inter-AS systems, while the other FIAs continue to rely on BGP. Most FIAs continue to work with any user-determined intra-AS algorithm, although MobilityFirst and NEBULA introduce their own.  NewIP expands intra-AS routing with additional features and capabilities. \textit{Multi-path routing} is the ability for flows to share multiple alternative paths through a network simultaneously, which provides enhancements to performance, security, and connection stability.  Every FIA supports multi-path routing to some degree, with NDN providing it in the most advanced form \cite{DBLP:journals/iotj/ZhangAYBA20}.

\mypara{Surveillance and censorship implications.} Censors may exploit content-centric networks using the in-network caching, data names and signatures, or file sharing to enhance their capabilities.  The similarity between DNS and  FIA-based name resolution services brings with it a large set of existing surveillance and censorship methods. The presence of additional information in packet headers (i.e., route, service) allows adversaries to create fine-grained censorship policies that can better detect evasion, re-direct packets, or block information sources.  The additional routing features generally do a good job of preventing existing TCP/IP-style censorship methods, however they introduce the potential for new attacks such as SCION path interposition or XIA route manipulation.

\section{Surveillance \& Censorship on FIA\MakeLowercase{s}}
\label{sec:censorship}
This section describes existing Internet surveillance and censorship techniques and we describe results how its effectiveness may change with the adoption of different FIAs. 

\subsection{Packet Inspection}
In deep packet inspection (DPI), the payload and headers of a packet are compared with a set of pre-defined patterns to identify similarities (e.g., the destination address or TCP port).  When a network packet matches the established criteria, the adversary can take some action (e.g., re-routing, dropping, or modifying the packet).  

\label{sec:features}
\mypara{Implications for surveillance and censorship.} 
DPI is frequently used by state-level adversaries to passively surveil internet usage in their country. For instance, Bourdillion \textit{et al.} ~\cite{DBLP:journals/clsr/Stalla-Bourdillon14a} described how mass surveillance using deep packet inspection is spreading around the world, becoming normalized and earning legal legitimacy. 
Furthermore, numerous experiments have been conducted searching for the presence of DPI-based censorship activity~\cite{DBLP:journals/popets/EnsafiWMC15,DBLP:journals/cn/AcetoP15,DBLP:conf/imc/YadavSGSC18}. Recently, Master and Garman~\cite{Master2023a} provided a more comprehensive outlook on how DPI mechanisms have been used by state-level censors across the world to issue blocking decisions.

\label{sec:collect}
\mypara{Impact on FIAs.} 
We now describe how DPI can be applied to FIAs' packets, comprising not only the analysis of packets' addresses and application-related content, but also of new header information like routing and service-level information.

\myparait{Address.}  Address-based filtering would work simply and effectively in SCION, NewIP, and NEBULA, as they all contain specific source and destination address information in the packet header.  In NDN, interest packets do not contain any reference to the specific source or destination host that a packet is travelling towards.  In XIA, source and destination addresses are required, however they can be partially encrypted to only specify an entire network.  Filtering traffic using address-only information in NDN and XIA would be highly ineffective and likely cause collateral damage \cite{DBLP:journals/ccr/X12}. In other FIAs like MobilityFirst, a single packet can have multiple destination addresses that can change mid-route, meaning that destination address filtering may also be largely ineffective. 

\myparait{Content.} The content-centric FIAs (i.e., NDN, MobilityFirst, and XIA) all contain a representation of the data contained within the packet payload as part of the destination address.  
In NDN, content is identified by its data name; in MobilityFirst, by the network address; and in XIA, by the content ID.  These addresses may prove valuable for adversaries engaging in censorship and surveillance as they can filter specific pieces of content, regardless from where it is obtained. In NDN and XIA the complete content address may be partially encrypted to protect against censorship and surveillance. In NewIP, packet contents may also be encrypted in separate chunks to aid with contract enforcement. 

\myparait{Route.} Some FIAs include routing information in packet headers, which can give rise to the development of new traffic filtering techniques based on the inspection of routing data.  
For instance, in SCION and XIA, route information is embedded into packet headers -- SCION contains the path of visited ASes within packet headers; XIA packets may contain a record of the entire path taken, specifying individual hops; NEBULA contains the entire cryptographically verified path within the header. These FIAs' packets can provide a finer granularity of information to adversaries w.r.t. TCP/IP (where only the source and destination of a packet is known).  

\myparait{Service.} The packet header structure of MobilityFirst, XIA, NewIP, and NEBULA supports the inclusion of service-layer information. However, this may mean that service information can be used to perform packet filtering. For instance, both MobilityFirst and NewIP contain plaintext service information within packet headers. Other FIAs like NEBULA and XIA mitigate this issue by supporting encrypted service abstraction layers between the network and transport layer, or supporting encrypted service IDs, respectively.

\begin{takeaway} 
\takeawaytitle{}
Many FIAs embed extra information and functionality in their network-layer packet design. This additional information can enable adversaries to improve DPI-based surveillance and create more detailed filtering policies.
\end{takeaway}

\subsection{Manipulation of Name Resolution}
\label{sec:dns}

The hierarchical DNS system relies on different servers located around the world. Users that need to resolve a hostname into an IP address typically query DNS servers owned by their local ISP (or organization).  However, there is no central body that directly governs the operation of low-level name servers  (i.e., the results they return) or that ensures the returned DNS responses will correctly reach back to the requesting client. Adversaries in control of DNS name servers can not only gather information about which websites clients wish to visit, but also manipulate DNS replies to redirect clients to different destination servers ~\cite{datta2022hannibal}.

\mypara{Implications for surveillance and censorship.} DNS provides an outlet for state-level adversaries to passively collect valuable intelligence about how their network is used by citizens, despite location and IP address changes~\cite{DBLP:journals/comsur/Schmid21}. For example, Snowden revealed large-scale DNS censorship efforts being undertaken by the NSA~\cite{grothoff2017nsa}, and Liu \textit{et al.}~\cite{DBLP:conf/anrw/LiuLDLLH019}  found that approximately 27\% of DNS requests from China to Google's servers were saved for analysis.

State-level censors can also exert their influence over the DNS name servers controlled by domestic ISPs, forcing them to block the resolution of specific blocklisted domain names~\cite{DBLP:conf/imc/YadavSGSC18,DBLP:conf/sp/NiakiCWHRCG20}, or inject fake DNS replies when users reach out to non-domestic DNS servers~\cite{DBLP:conf/srds/HouserH0LCW21}. In both cases, the adversary can either return no information to users, or provide them with specific details about why a specific access was disallowed~\cite{DBLP:conf/uss/VerkampG12}. 

\label{sec:newdns}
\mypara{Impact on FIAs.} 
With the exception of NDN and MobilityFirst, the FIAs we studied still leverage the current Domain Name System (DNS) to provide address--to--hostname mappings, thus maintaining the status quo with regards to censorship and surveillance efforts. In NEBULA, however, DNS hijacking can potentially be made more difficult for an adversary, as DNS client resolutions are augmented with policy enforcement.  
MobilityFirst takes a distributed approach for handling name resolution, meaning a censor would have limited ability to access  or manipulate the records stored by the name resolution service.  However, DNS hijacking is still a present threat. In NDN, if names are resolved with a deterministic algorithm that could be run locally, no queries or network interaction would be involved.  In such a setting, an adversary would be unable to engage in any form of DNS-based censorship. However, since exclusively local name resolution is impractical, some solutions that support using the existing DNS infrastructure for NDN name resolution~\cite{DBLP:conf/acmicn/TehraniOSSW19} have been proposed. NDN-DNS follows a similar structure to traditional DNS, so poisoning and hijacking attacks are possible.

Theoretically, widespread implementation of DNSSEC or similar would severely limit or entirely prevent surveillance and censorship efforts focused on name resolution services~\cite{Bator_Przystasz_Serafin_2023}. However, DNSSEC's current iteration has faced several issues, including continued breaches of its purported defenses~\cite{DBLP:journals/comsur/Schmid21}, as well as an extremely slow adoption~\cite{Huston_2024}. While some work has proposed DNSSEC-like tools for NDN~\cite{DBLP:conf/ictc/ParkSK20, DBLP:conf/acmicn/TehraniOSSW19}, they are also exposed to similar security and operational challenges as TCP/IP's DNSSEC.

\begin{takeaway} 
\takeawaytitle{}
The majority of FIAs make no changes that prevent wide-scale DNS-based censorship efforts.  Although some FIAs prevent either DNS hijacking or poisoning, no design defends against both (other than NDN in very specific use-cases).
\end{takeaway}

\subsection{Traffic Analysis}
\label{subsec:traffic_analysis}

Traffic analysis is the process of examining a network's connection \textit{metadata} with the purpose of inferring information about the communications of users (e.g., websites they browse~\cite{DBLP:journals/popets/RahmanSMG020} or whom they speak to via instant-messaging~\cite{DBLP:conf/ndss/BahramaliHSGT20}), even when these communications are encrypted. This metadata may include a connection's data volume, packet sizes and arrival times, or identifiers that remain unencrypted, e.g., addresses, for routing purposes.  As an example for the following exposition, we focus on a prominent traffic analysis attack known as website fingerprinting~\cite{DBLP:conf/uss/HayesD16,DBLP:conf/wpes/WangG13}, where an adversary listens to the encrypted traffic produced by a user when accessing a website (e.g., sent through some encrypted tunnel like Tor), and attempts to identify which website the user visited. The adversary does so by comparing the metadata of the traffic exchanged by the user with a set of metadata previously collected by the adversary when using the same encrypted tunnel. 

\mypara{Implications for surveillance and censorship.} 
Traffic analysis attacks such as website fingerprinting~\cite{DBLP:conf/uss/CherubinJT22,DBLP:conf/sp/DengYLZLXXW23} can be leveraged by a state-level adversary both for surveillance (e.g., keep a list of which websites a user visits over an encrypted tunnel), and for censorship (e.g., block a user's Internet connection if they are found to visit a prohibited website).  In addition to website fingerprinting,  similar attacks have also been used to uncover the utilization of network covert channels that allow users to circumvent censorship~\cite{DBLP:conf/uss/Barradas0R18,DBLP:conf/ccs/WangDARS15,DBLP:conf/ccs/GeddesSH13}. These techniques have been used in other countries to assist with internet censorship efforts ~\cite{DBLP:journals/ieeesp/DixonRS16,DBLP:conf/uss/WuSSBAW0HLW23}. 

\mypara{Impact on FIAs.} To reason about the potential impact of fingerprinting attacks on FIAs, we consider four major categories of features used in website fingerprinting attacks, as identified by Wang and Goldberg~\cite{wang_goldberg}. 
We address them below.

\myparait{Packet length.} Packets in TCP/IP use fixed-length headers (i.e., 20 bytes for TCP), thus typically providing minimal information about the content of each packet. In contrast, a majority of FIAs' packet headers support variable lengths.  For example, NDN interest packet lengths can help to infer the content being requested (even while encrypted) in the same way that the length of packets carrying an HTTP GET request may be a valuable feature for website fingerprinting.  In SCION, XIA, and NEBULA, packets contain variable length headers depending on the length of the route the packet takes. NewIP packets' length is highly variable due to the presence of contracts in the header (albeit these contracts may already be used to infer what type of data is being exchanged).

\myparait{Packet length frequency.}
These distributions would likely remain a useful feature for characterizing the traffic produced by all the considered FIAs. Any changes of the internet design concepts would still be limited by hardware transmission capacity (i.e., Ethernet's maximum transmission unit of 1500 bytes), which means the overall distribution of packet lengths will still provide useful information for launching fingerprinting attempts.

\myparait{Packet ordering.}
Depending on how each FIA configures two-way communication, packet ordering characteristics can be impacted. In particular, NDN uses a content-centric design where each outgoing interest must be strictly matched by incoming data packets, in a pull-model fashion.  Even if the outgoing interests and corresponding incoming data packets do not follow a strict one-to-one relationship, network observers can match data packets with interests using the embedded name.  The other content-centric networks do not utilize an interest/data packet relationship in the same way and thus do not share such a potential vulnerability.

\myparait{Packet timing.}
The timing between consecutive packets as a feature for performing accurate traffic analysis is highly affected by each FIA's unique approach to routing.  In the content-centric designs (NDN, MobilityFirst, and XIA), inter-packet timing is inconsistent when packet arrival times shift due to cache retention and replacement algorithms. Another design choice that may impact the usefulness of packet timing is the capability to perform multi-path routing. For instance, in MobilityFirst and XIA, packet routes can be changed mid-transit, leading to variability in the time a packet takes to reach a machine responsible for collecting traffic for further analysis.  In SCION, XIA, and NEBULA, hosts are also able to choose the path for their packets ahead of time, making it trivial to manipulate travel times. In NewIP, contract specifications may demand processing delays, leading to highly variable timing.

\begin{takeaway} 
\takeawaytitle{}
There is little evidence to suggest that the changes made to packets by FIAs inherently resist traffic analysis attacks. More empirical research would be required to assess FIAs' susceptibility to traffic analysis attacks.
\end{takeaway}

\subsection{Packet Manipulation}

Adversaries might attempt to modify packet contents or change the way they move through the network~\cite{DBLP:conf/ws/MeyerW04}. 

\mypara{Implications for surveillance and censorship.} The ability to modify packets that are en-route is a valuable tool for state-level adversaries looking to extend surveillance and censorship capabilities. First, an adversary could manipulate the route listed in a packet header to re-route it towards a DPI-capable middlebox, or to completely disrupt communication between two hosts (e.g., by explicitly dropping packets). When sitting between the communication of two hosts, adversaries can also change packet contents', e.g., replacing the contents of a DNS reply, or inject packets into endpoints' communications, e.g., spoofing the address of a server and prompt a client within the censored region to tear down an ongoing connection via fake connection reset packets~\cite{DBLP:conf/pet/ClaytonMW06}.

\mypara{Impact on FIAs.} 
We now describe how adversaries can manipulate or inject packets in different FIAs' network flows.

\myparait{Route manipulation.} 
In SCION and NEBULA, modifying packets' path directly has been specifically defended against with the use of cryptographic next-hop verifiers.  In MobilityFirst and NewIP, packet headers can be indirectly manipulated by an adversary to modify its path. In MobilityFirst, the adversary can strategically add or removing the interfaces for a specific GUID, while in NewIP, packet forwarding rules can be adjusted dynamically using the contract headers. Finally, in XIA, there are no specific protections in place to prevent the adversary from changing a packet's route.

\myparait{Packet injection.} 
No FIAs make any changes that would prevent packets from being injected or dropped from a given network flow. However, tracing a specific flow is decidedly more difficult in content-centric networks due to inconsistent address information (e.g., MobilityFirst's dynamic addresses) and packet aggregation~\cite{DBLP:conf/acmicn/KhoussiPBB17}.  Many FIAs make efforts to prevent replay~\cite{DBLP:conf/csfw/Syverson94}, connection reset~\cite{DBLP:reference/crypt/Weaver11}, and other integrity-violation attacks~\cite{SCION_book}. 

\begin{takeaway} 
\takeawaytitle{}
Packet manipulation is markedly more difficult in several FIAs, as integrity checks and defenses were carefully considered at their design stages.
\end{takeaway}

\subsection{Attacks on Network Caching Infrastructure}
Network caches have become a fundamental component of the Internet infrastructure, allowing content providers in the web to serve content faster while reducing bandwidth loads and costs in enterprise networks~\cite{DBLP:conf/ccs/NguyenIF19,DBLP:conf/imc/MouraHSH19}. In the current Internet, caching can be implemented at different layers, including caching proxies (e.g., Squid~\cite{DBLP:conf/sigmetrics/RousskovS98}) or origin web servers (e.g., NGINX~\cite{nginx}).  
From a privacy perspective, however, the presence of caches may allow adversaries to deduce information about the data accessed by users~\cite{DBLP:conf/imc/RandallLAPVSS20,DBLP:conf/uss/MirheidariAOCK020}, or prevent users from accessing the intended data~\cite{DBLP:conf/ccs/NguyenIF19}. Outside of network caches, Chaabane \textit{et al.}~\cite{DBLP:journals/ccr/AbdelberiCKU13} propose two additional attack vectors for enabling adversaries to engage in censorship and surveillance over NDN.  First, the human-readable addresses used by NDN have the potential to reveal information about the underlying content, since they are not abstract like in XIA or MobilityFirst.  Second, NDN signatures used to sign data might reveal sensitive information about the original data producer.

\mypara{Implications for surveillance and censorship.} 
We examine five main attack vectors involving network data caches that a state-level adversary could use to infer private information over FIAs. NDN is the primary FIA where these attacks have been studied and have been considered feasible to some extent. 

\myparait{Cache enumeration.} This attack enables an adversary to identify contents stored in a network cache, potentially linking them to the requesting users. If a caching router is under an adversary's control, it may scan for specific data names on a blocklist to identify users requesting prohibited content. Adversaries can also list the contents of caches outside their jurisdiction by repeatedly requesting random or deterministically generated content from those caches.~\cite{DBLP:conf/noms/CompagnoCLTV20}.

\myparait{Cache pollution.} In this attack, an adversary floods a router’s cache with requests for obscure data displacing previously cached legitimate content ~\cite{DBLP:conf/softcom/HidouriHHTM21}. Pollution attacks can be employed around hosts suspected of serving prohibited content, causing severe slowdowns, akin to throttle-based censorship~\cite{DBLP:conf/imc/XueRSEVJWBE21}.

\myparait{Content poisoning.} In this attack, adversaries may place content with a valid name and fake content or signatures into a router cache~\cite{DBLP:conf/im/NguyenMDCC17}.  When a user requests poisoned content, the fake version is served as opposed to being forwarded to the original source.

\myparait{Cache timing.} An adversary can determine if some content has been recently accessed by a particular target ~\cite{DBLP:conf/ccnc/TsudikUW16}. First, the adversary determines the retrieval time for cached and non-cached content from a nearby router by sending out interests.  If the response time matches their recorded time for a cached interest, it means the target user has recently loaded that content.

\myparait{Conversation cloning.} In this attack, an adversary attempts to replicate a flow in the cache.  First, the attacker discovers the specific pattern with which content names are generated for some flow (e.g., a voice-over-CCN with IDs \texttt{/users/alice/1,2,3}).  Then, the attacker can predict future data packets by sending interests matching the discovered pattern and gain access to the packet contents.

\mypara{Impact on FIAs.} The long-lived content caches in NDN, MobilityFirst, and XIA makes cache enumeration, timing, and conversation cloning possible. In particular, NDN automatic prefix matching allows an attacker to systematically map out the content of nearby router caches ~\cite{DBLP:journals/ccr/LauingerLRSBK12}. In turn, MobilityFirst and XIA do not support automatic prefix matching; to perform enumeration attacks on these architectures, an adversary must narrow their records to specific websites or content.   Cache poisoning operates in a similar fashion as DNS cache poisoning is performed in the existing Internet. Cache poisoning is not possible in MobilityFirst or XIA since data contents make use of self-certifying names. In turn, cache pollution attacks are possible in NDN, MobilityFirst, and XIA.   

\begin{takeaway} 
\takeawaytitle{}
The presence of router data caches in content centric FIA networks opens up a broad surface for surveillance and censorship efforts to take place. 
\end{takeaway}

\subsection{Summary}

Table~\ref{tab:techniques} depicts an overview of the applicability of different censorship and surveillance techniques to the FIAs we consider. All the FIAs are susceptible to DNS-based censorship, a widely employed technique in today's TCP/IP censorship apparatus.  In content-centric networks, caching is likely to become a key focus for surveillance and censorship efforts.  Packet inspection is more effective in FIAs that implement complex routing systems but requires greater efforts from adversaries, thus presenting only a moderate threat. Fingerprinting is still possible, although identifying specific flows may prove difficult in content-centric networks. Packet manipulation is effective in FIAs as many integrity protection mechanisms have been added. Packet manipulation and fingerprinting may be less severe compared with the aforementioned approaches because they require comprehensive monitoring systems and considerable processing power to be effective (e.g., to adequately train and deploy traffic classification algorithms).

\section{Privacy Enhancements for FIAs}
\label{sec:defense}
The continuous cycle of building tools for evading increasingly sophisticated Internet censorship and surveillance apparatuses is typically referred to as an arms race~\cite{DBLP:conf/indocrypt/Dingledine11}. This section describes a set of prominent privacy-enhancing technologies introduced to address network surveillance and censorship concerns, and which have seen practical deployments in TCP/IP networks. We aim to capture important classes of privacy-enhancing technologies that have gained traction within the research community, and which fostered ongoing improvements. Building on this description, we assess how these techniques may translate to the FIA context.

\mypara{Onion routing.} Currently, NEBULA, SCION, and NDN are able to support onion routing protocols that provide anonymity guarantees for clients and content providers on the network. NEBULA can integrate TorIP \cite{DBLP:conf/hotnets/LiuHKA11} by default. SCION, MobilityFirst, and XIA have their own tailored anonymity protocols, which we detail below. Content-centric networks do not allow for traditional onion routing, but alternative anonymity systems have been devised~\cite{DBLP:conf/sp/HsiaoKPYNGM12}.

Past research has looked into different approaches to provide Tor-like anonymity on NDN. In ANDaNA~\cite{DBLP:conf/ndss/DiBenedettoGTU12}, each router participating in the network takes an incoming interest, encrypts it as the payload of a new interest packet with the same name and sends it forward. In this way, the signatures of interest and data packets can only be seen by routers one link before and after along the packet path. AC3N~\cite{DBLP:conf/ccnc/TsudikUW16} improves the performance and throughput of ANDaNA by using MACs to ensure packet integrity within an ANDaNA network. Seo \textit{et al.}~\cite{6994394} also propose a Tor-like anonymity system for content-centric networks. HORNET \cite{DBLP:conf/ccs/ChenABDP15} is a Tor-like anonymization protocol designed for integration within SCION that uses shared symmetric keys between routers along the pre-defined SCION path to hide the complete path from intermediate nodes and prevent path manipulation.  TARANET \cite{DBLP:conf/eurosp/0013APBDT18} adds mixing, traffic shaping, and packet splitting to HORNET as a defense against traffic analysis attacks. Other schemes also aim to cryptographically protect the entire SCION path \cite{DBLP:conf/europ4/SchulzWH23}.

\begin{table}[t!]
\caption{Summary of censorship and surveillance techniques and their effectiveness in each of the studied FIAs.} 
\label{tab:techniques}
\vspace{-0.3cm}
\resizebox{\linewidth}{!}{
\begin{tabular}{lccccccc}
\hline
\textbf{Technique} & \textbf{NDN} & \textbf{SCION} & \textbf{MF} & \textbf{XIA} & \textbf{NewIP} & \textbf{NEBULA} & \textbf{IP} \\ \hline
\multicolumn{8}{c}{\textbf{Packet Inspection}} \\
Address & \emptycirc & \fullcirc & \halfcirc & \halfcirc & \fullcirc & \fullcirc & \fullcirc \\
Content & \fullcirc & — & \fullcirc & \fullcirc & — & — & — \\
Service & \emptycirc & \emptycirc & \fullcirc & \halfcirc & \halfcirc & \halfcirc & \emptycirc \\
Route & \emptycirc & \halfcirc & \emptycirc & \halfcirc & \emptycirc & \fullcirc & \emptycirc \\
Hostname & \emptycirc & \fullcirc & \halfcirc & \fullcirc & \fullcirc & \fullcirc & \fullcirc \\
URL & \fullcirc & \fullcirc & \fullcirc & \fullcirc & \fullcirc & \fullcirc & \fullcirc \\ \hline
\multicolumn{8}{c}{\textbf{Name Resolution Manipulation}} \\
Poisoning & \halfcirc & \fullcirc & \halfcirc & \fullcirc & \fullcirc & \fullcirc & \fullcirc \\
Hijacking & \fullcirc & \fullcirc & \fullcirc & \fullcirc & \fullcirc & \fullcirc & \fullcirc \\ \hline
\multicolumn{8}{c}{\textbf{Traffic Analysis}} \\
Metadata & \halfcirc & \halfcirc & \halfcirc & \halfcirc & \halfcirc & \halfcirc & \halfcirc \\ \hline
\multicolumn{8}{c}{\textbf{Packet Manipulation}} \\
Manipulate Route & \emptycirc & \fullcirc & \emptycirc & \fullcirc & \halfcirc & \emptycirc & \emptycirc \\
Drop Packet & \fullcirc & \fullcirc & \fullcirc & \fullcirc & \fullcirc & \fullcirc & \fullcirc \\
Manipulate Router & \emptycirc & \emptycirc & \emptycirc & \halfcirc & \fullcirc & \emptycirc & \emptycirc \\
Spoof Address & \fullcirc & \fullcirc & \fullcirc & \fullcirc & \fullcirc & \fullcirc & \fullcirc \\ \hline
\multicolumn{8}{c}{\textbf{Network Caching Infrastructure Attacks}} \\
Enumeration & \fullcirc & — & \halfcirc & \halfcirc & — & — & — \\
Timing Attack & \fullcirc & — & \fullcirc & \fullcirc & — & — & — \\
Cloning & \fullcirc & — & \fullcirc & \fullcirc & — & — & — \\
Poisoning & \fullcirc & — & \emptycirc & \emptycirc & — & — & — \\
Pollution & \fullcirc & — & \fullcirc & \fullcirc & — & — & — \\ \hline
\multicolumn{8}{l}{\fullcirc[0.7ex] = Technique Possible. \halfcirc[0.7ex] = Limited Possibility. \emptycirc[0.7ex] = Not Possible. — = N/A.} \\ \hline
\end{tabular}
}
\vspace{-0.1cm}
\end{table}

\mypara{Decoy routing.} From a purely architectural standpoint, decoy routing systems should be feasible to implement on all the host-based FIAs we have discussed. In fact, such schemes have already been developed for NDN~\cite{DBLP:conf/globecom/MozaffariHV19}, and related systems such as Harpocrates~\cite{DBLP:conf/sacmat/AzadTMM22} similarly leverage proxy nodes to hide sources of data from censors in NDN.  Moreover, the general flexibility of addresses in MobilityFirst and routes in XIA suggests that decoy routing may be much easier to implement when compared with TCP/IP. SCION's public path listings can provide clients a guarantee that their packet will reach and be processed by a given decoy router, possibly deterring adversaries' attempts at avoiding paths containing such routers~\cite{DBLP:conf/ccs/SchuchardGTH12}. In turn, we expect NEBULA's cryptographic route guarantees and consent requirements to complicate decoy routing since the paths packets must follow after reaching a decoy router must somehow be pre-approved and verified.

\mypara{Application-layer covert channels.} Covert channels have been developed at the application layer to allow clients located within a censored region to secretly access blocked contents or destinations~\cite{DBLP:conf/sp/TschantzAAP16}. In this setting, a client  initiates a connection with a server located in the free Internet region over some application  that is allowed to cross a censor's border. Resorting to different techniques~\cite{DBLP:conf/ccs/Barradas0RN20, DBLP:conf/asiaccs/SharmaGC21,DBLP:conf/sp/SunS23,DBLP:conf/uss/RosenPM21}, the application's messages can be manipulated to transfer potentially blocked content instead of legitimate application data.  Covert channels can be implemented on the application-layer in each of the FIAs.  However, ensuring that one-to-one covert channels can be established over FIA protocols might prove challenging. For instance, in IP-multicast protocols (i.e., the methods used in MobilityFirst and XIA), the current service type may be listed within packet headers, making it difficult to generate one-to-one traffic over multicast protocols.

\mypara{Bootstrapping.} A problem shared between many anonymity and censorship circumvention tools is the method for a user to securely exchange cryptographic keys to bootstrap her activity in the system.  Some tools \cite{DBLP:conf/osdi/LazarZ16} help with bootstrapping anonymous communication in TCP/IP, but no network-layer solution exists and many censorship-evasion techniques rely on out-of-band key exchange~\cite{DBLP:journals/sncs/LatvalaSA20}.  Interestingly, NDN and XIA provide some built-in solutions to the bootstrapping problem.  In NDN, a deterministic name generation algorithm can run on hosts and routers.  In XIA and NDN, interests can be sent with missing components, leaving intermediate routers to fill-in the blanks. A client might direct their packets to a large network and allow routers within that network to determine the correct endpoint for a packet.

\mypara{Blockchain.} Recently developed privacy enhancing technologies have integrated blockchain as a mechanism to enhance privacy and limit surveillance and censorship efforts. In particular, InterPlanetary File Sharing~\cite{DBLP:journals/corr/Benet14} (IPFS) is a peer-to-peer distributed file sharing system with the goal of spreading internet resources more equitably, as the majority of current TCP/IP web traffic originates from a few particular organizations. IPFS (and similar) blockchain-based file sharing systems are not immune to censorship~\cite{DBLP:conf/ndss/SridharAKGPPRK24, DBLP:conf/ssci/AcharjamayumPD18}. Meanwhile, IPFS' peer-to-peer nature makes surveillance arguably easy, as users leave a public record of the data they wish to access and when it is downloaded~\cite{DBLP:conf/sigcomm/TrautweinRTCSSG22}. Other censorship-resistant communication tools have also been developed using public blockchains~\cite{DBLP:conf/eurosp/TiemannBEL23}, permissionless cryptocurrencies~\cite{DBLP:journals/popets/MinaeiMK20}, or Satoshi (i.e., Bitcoin) block chains~\cite{DBLP:journals/popets/RecabarrenC22}. In NDN, tools for using blockchain~\cite{DBLP:journals/jnca/ThaiKBK22} and applications to provide identifier management~\cite{DBLP:journals/access/YangCS19}, cache protection~\cite{DBLP:journals/grid/LeiFZLDHWX20}, and key management~\cite{8605993} have been proposed.  A distributed file sharing system targeted at SCION (and based in blockchain) has also been developed~\cite{ryll2018development}. NewIP has been the target of proposals to use blockchain for name resolution services, although these are likely to be more vulnerable to censorship and surveillance than DNS itself~\cite{wu20245g, caeiro2021technical}. MobilityFirst and XIA's content-centric design makes them well-suited for potential integrations with blockchain and IPFS-like technologies, although none have been developed.

\mypara{Content anonymity.} Some research has considered different approaches to anonymizing content-centric data requests (without attempting to hide the host that sent an interest). Tourani \textit{et al.}~\cite{DBLP:conf/acmicn/TouraniMKOM15} propose partially encoding NDN interest packets using multiple Huffman tables.  The data name prefix would direct packets to an anonymization network followed by the precise encoded identifier. Censors can only see that a packet is directed towards a network that may be too large to censor in its entirety.  Feng \textit{et al.}~\cite{Feng2015/06} also propose a similar encoding scheme.  In turn, Fotiou \textit{et al.}~\cite{DBLP:journals/scn/FotiouTMKP14} suggest using homomorphic encryption for concealing the destination of NDN interest packets; data names are hierarchically organized such that a censor can only interpret the initial packet destination.  For example, a packet would list \texttt{/ufia/ad729fe/cab609a/...} as the interest.  Once a router within \texttt{ufia} receives the packet, they can decrypt the next segment and forward it as required. PrivICN~\cite{DBLP:journals/cn/BernardiniMAC19} is a system that expands on existing work by offering full protection to data producers and consumers.

An alternative approach to anonymizing content in NDN is to duplicate data, where a user requests content from a trusted source who layers the original file with their own encryption and signature. However, identifying duplicated content is a key operation for the optimization of large-scale NDN deployments, which may conflict with the deployment of safeguards against censorship. Dulal \textit{et al.}~\cite{DBLP:conf/acmicn/DulalW23} propose an algorithm to automatically identify and resolve copies of the same content being cached at the same location.

\mypara{Cache protection.} Extensive research has investigated defenses against the array of data cache attacks present in NDN. Qu \textit{et al.} and \cite{DBLP:journals/ton/QuLQSYH22} propose using a blockchain to prevent pollution attacks, while Lei \textit{et al.} \cite{DBLP:journals/grid/LeiFZLDHWX20} suggest that blockchains may be used as a countermeasure for poisoning.  Hyeonseung \textit{et al.} \cite{DBLP:journals/itiis/Im020} provide a thorough analysis of cache poisoning attacks and defenses.  Multiple pollution mitigation techniques use statistical analysis to determine the probability of an ongoing attack, but these can be evaded by name encryption and obfuscation techniques~\cite{DBLP:journals/jcst/KumarSAS19}. \'{A}cs \textit{et al.}~\cite{DBLP:journals/tdsc/AcsCGGTW19} also propose a defense against cache timing attacks in NDN by artificially delaying specific content requests.  Incentive-based caching~\cite{DBLP:journals/access/NdikumanaTHNHH18} as a mechanism for reducing costs for large-scale NDN deployments may also serve as a deterrent for censors that extensively request content from or test particular routers extensively.

\mypara{Summary.} Table ~\ref{tab:defense} depicts an overview of the ongoing deployment of the different surveillance and censorship defenses addressed above. NDN and SCION have seen the most research into privacy enhancing techniques for defending against internet surveillance and censorship, while, some FIAs allow for trust bootstrapping.

\begin{table}[t!]
\centering
\caption{Application and implementation of prominent privacy enhancing techniques for FIAs.}  
\vspace{-0.3cm}
\label{tab:defense}
\resizebox{\linewidth}{!}{
\begin{tabular}{@{}lccccccc@{}}
\toprule
\textbf{Technique} & \textbf{NDN} & \textbf{SCION} & \textbf{MF} & \textbf{XIA} & \textbf{NewIP} & \textbf{NEBULA} & \textbf{IP} \\ \midrule
Onion Routing & \implemented & \implemented & \implemented & \implemented & \fullcirc & \implemented & \implemented \\
Decoy Routing & \implemented & \fullcirc & \fullcirc & \fullcirc & \fullcirc & \emptycirc & \implemented \\
Covert Channel & \implemented & \implemented & \halfcirc & \implemented & \halfcirc & \fullcirc & \implemented \\
Bootstrapping & \implemented & \emptycirc & \emptycirc & \fullcirc & \emptycirc & \emptycirc & \emptycirc \\
Blockchain & \implemented & \implemented & \fullcirc & \fullcirc & \implemented & \halfcirc & \implemented \\
Content Anonymity & \implemented & — & \implemented & \emptycirc & — & — & — \\
Cache Protection & \implemented & — & \fullcirc & \fullcirc & — & — & — \\ \midrule
\multicolumn{8}{l}{\implemented = Implemented. \fullcirc[0.75ex] = Possible. \halfcirc[0.75ex] = Limited Possibility. \emptycirc[0.75ex] = Not Possible. — = N/A} \\ \bottomrule
\end{tabular}
}
\vspace{-0.1cm}
\end{table}
\section{Discussion \& Open Challenges}
\label{sec:futurework}
Informed by our previous analysis, this section discusses potential directions for future work tied to the empirical assessment of FIAs' susceptibility to different classes of privacy-invasive attacks aimed at enforcing network surveillance and censorship policies and the development of privacy-enhancing technologies tailored for FIAs.

\mypara{FIA testbeds.} First, we highlight the current dearth of easy to deploy experimental FIA testbeds that would allow practitioners to perform practical evaluations of FIAs' defensive capabilities in the above settings~\cite{DBLP:journals/ccr/MastorakisAZ17, DBLP:conf/conext/KrahenbuhlTGKPH21}. The existing tools face some important challenges that slows FIA research. First, most FIA testbeds are developed for a closed world or locally created network.  We suggest that FIA software should provide a solution to integrate with the existing internet (using one of the many systems developed~\cite{electronics12071723}), to provide researchers with improved data collection capabilities. Second, FIA testbeds are developed under different languages and low-level libraries~\cite{DBLP:journals/tcom/TanFLJZY20, DBLP:journals/ccr/MastorakisAZ17, SCION_book}, resulting in that the experimentation with new or upgraded FIA security-focused architectural elements (which may even be applicable amongst multiple FIAs) requires extensive custom development for different  testbeds. Finally, research into the interoperability of different FIAs (and security thereof) is  an aspect that remains overlooked, with most existing research targeting the integration of a given FIA with TCP/IP.

\mypara{Resisting fingerprinting and packet inspection.} A wide range of traffic analysis techniques have been developed to conduct effective packet and flow inspection in the current Internet. However, it remains uncertain whether the most valuable attributes for IP-based fingerprinting would also hold a similar relevance when considering network traffic exchanged via FIA-based networks. Some defenses against traffic analysis have been proposed for FIAs (e.g., AC3N and HORNET), but it is unclear whether the effectiveness of such tools can be upheld under a closer scrutiny from the research community. For instance, see the multiple attacks and improvements on Tor experienced over the last decade~\cite{DBLP:journals/comsur/KarunanayakeAMI21}, as well as recently documented flaws~\cite{DBLP:conf/sp/KuhnBS20} on anonymity solutions (such as HORNET~\cite{DBLP:conf/ccs/ChenABDP15}) that envision a deployment within FIAs. 

Based on current research in IP and FIAs, we identify a set of thrusts that can help gauge the effectiveness of FIA defenses against packet inspection. Specifically, these include assessments of the ability of adversaries to: a) identify network flows and their sources/destinations; b) deduce the purpose of an encrypted flow through analysis of FIA traffic metadata, and; c) identify FIA packet attributes that can be exploited to enact filtering policies. Hiding one address on a packet alone cannot stop adversaries from identifying a packet's destination network or interface in some FIAs.

\mypara{Resisting censorship.} There have been a wide array of privacy enhancing tools constructed to defend against TCP/IP-based censorship and surveillance. While some of these tools have been adapted to the FIA context (\S\ref{sec:defense}), many have no equivalent implementation or discussion surrounding their value to FIAs. We posit that one important reason for this fact is that some of these tools require large support bases and/or infrastructures which are not yet available in current FIA deployments~\cite{ DBLP:conf/uss/XueARDE24}. 

We believe that the industry is well-positioned to help overcome the above obstacles (even if indirectly) by investing in large-scale FIA deployments (e.g.,~\cite{anapaya}).  Indeed, prioritizing advances on the technical readiness of FIAs will not only foster their early adoption and the potential creation of innovative network services to users, but also enable the research community to refine and scale measurements and/or anti-censorship tools on FIAs more effectively, departing from simulation-led experiments (e.g.,~\cite{DBLP:journals/corr/abs-2403-09447}). Provided with such capabilities, the research community should be better equipped to re-create/extend existing censorship-resistant communication tools into the architecture space of FIAs.

\mypara{Resisting caching and name resolution attacks.} Multiple studies have investigated potential flaws and attack vectors for the data caching mechanisms used in content-centric networks.  Complementary research has proposed strategies to mitigate the effects of specific attacks on caching and proposed broader defense mechanisms to prevent abuse.  One aspect of FIAs that remains unexplored through the lens of censorship and surveillance pertains to the interplay between name resolution services and caching. As it stands, a large fraction of current internet censorship mechanisms are triggered at DNS servers and their caches, a trend which is likely to persist in FIA deployments. While security extensions like DNSSEC~\cite{DBLP:journals/rfc/rfc4033} have been devised to counteract cache poisoning and pollution in the current internet, the deployment of this protection faces slow adoption. An interesting direction for future work would be to undertake a comprehensive analysis of security and privacy aspects concerning emerging name resolution services (such as NDNS and GNRS), alongside practical assessments of how attacks on in-network caching might impact these services.

\mypara{Resisting packet manipulation.} 
Previous studies have explored how adversaries might manipulate Internet packets' paths or headers. Several defenses and safeguards have been incorporated within FIAs to counter these exploits, however few works have quantitatively assessed the efficacy of these defenses. Follow-up research could uncover novel vulnerabilities within FIAs~\cite{DBLP:journals/corr/abs-2405-06074}, leading to important enhancements on packet manipulation safeguards.  We believe further research should analyze how packet paths are explicitly created and followed in each FIA, with the intent of understanding adversaries' ability to: a) modify a packet's route to enact filtering policies or disallow it from leaving the region; b) drop packets within the network without the source or destination realizing it; c) manipulate the network's response to a packet (particularly in NewIP) and assess the practical implications of a successful execution, and; d) spoof network addresses.  

\mypara{Resisting Internet shutdowns.} Adversaries with control over a country's network infrastructure can enforce \textit{shutdowns}, i.e., block all access to and from any devices within their jurisdiction, or enact partial regional shutdowns~\cite{DBLP:conf/amcis/LeberknightR18}. Dainiotti \textit{et al.} ~\cite{DBLP:journals/ton/DainottiSACCRP14} reported on two main approaches used to perform these shutdowns: disrupting BGP connectivity, e.g., by halting the proper announcement of BGP routes, and performing extremely restrictive packet filtering.

Our analysis suggests that the above shutdown mechanisms may be applicable to all the FIAs discussed. Note that NDN, MobilityFirst, NewIP, and NEBULA use a BGP style routing protocol that would be equally affected by current disruption techniques. We believe further research should concentrate on: a) improving the robustness of inter-domain routing algorithms in existing/future FIA designs, towards developing mechanisms that might deter adversaries from performing disconnections (e.g., by introducing accountability mechanisms~\cite{accountability}), and; b) develop capabilities for FIA users to recover connectivity when subjected to shutdowns.

\section{Conclusion} In this paper, we analyse the different design elements of six prominent FIAs, pinpointing conceivable avenues of attack that could be exploited by powerful network adversaries to surveil and censor Internet communications. Our analysis encompasses past efforts in designing countermeasures against such practices, while shedding light on areas requiring further attention. FIAs introduce novel architectural elements aimed at improving Internet efficiency, scalability, and security. These enhancements disregard new protections against the efforts of powerful state-level network adversaries to enforce tight network monitoring and control policies. As a result, FIA designs can inadvertently facilitate the deployment of mass surveillance and censorship tools. 

\begin{acks} 
This work was partially supported by NSERC under grants DGECR-2023-00037 and CGS D-578769-2023, as well as the David R. Cheriton Chair in Software Systems.  We would also like to thank Adrian Perrig for valuable discussions about this work.
\end{acks}

\pagebreak

\bibliographystyle{ACM-Reference-Format}
\bibliography{refs}

\appendix

\section{The Internet Protocol (IP)}
\label{bg:ip}
The Internet Protocol (IP)~\cite{DBLP:journals/rfc/rfc791} is the network-layer communication protocol used in the current Internet that directs packets from a source to a destination based on the pair of addresses provided. IP consists of a simple design that provides best-effort service with no guarantees (i.e., packets may be lost, fragmented, or re-ordered).  More complex features (e.g., security, congestion control, reliable delivery) that are required for modern Internet functionality are provided by other protocols in different layers.

\mypara{Names and addresses.} 
IPv4 addresses are 32 bits long and organized according to the classless inter-domain routing standard~\cite{DBLP:journals/rfc/rfc4632}. Each address is split into a network-identifying prefix followed by a host identifier.  Organizations (e.g., ISPs) are assigned IP address blocks by the Internet Assigned Numbers Authority. In turn, the Domain Name System (DNS) \cite{DBLP:journals/rfc/rfc1035} serves as the phonebook of the Internet, and converts human-readable hostnames (e.g., \url{www.example.com}) into IP addresses (e.g., 192.168.0.1).  DNS name servers store mappings between hostnames and IP addresses, and respond to queries about their records. These servers are distributed around the world and organized hierarchically.

\mypara{Packet structure.} 
IPv4 packet headers~\cite{DBLP:journals/rfc/rfc791} contain 20 bytes (+ a 40-byte optional section).  Header fields include the source and destination address, its length, the payload length, checksum, packet fragmentation data, and upper-layer protocol information. 

\mypara{Routing.} 
In IP, routing is organized into a separate data and control plane.  The data plane encompasses low-level operations concerned with how routers process packets, and is focused on determining packets' correct outgoing links.  In turn, the control plane can be perceived as a high-level configuration that determines how packets should flow through a network. An autonomous system (AS) is a group of IP address prefixes owned by one or more network operators with a clearly defined routing policy. Control plane routing is thus divided into intra-AS and inter-AS.

The data plane consists of three main operations: input processing, switching, and output processing.  A router determines the output link for an incoming packet by comparing header information (e.g., destination IP address) with entries in its forwarding information base (FIB) using the longest prefix matching rule. The router's switching fabric then quickly sends the packet to an outgoing link. If packets arrive at the output of a router above the line rate, they are stored in a queue. The network operator's routing policy determines which packets are forwarded first.

In the control plane, intra-AS routing primarily focuses on finding the most efficient path between two nodes within an AS. There are numerous different approaches to intra-AS routing such as RIP~\cite{DBLP:journals/rfc/rfc2453} and OSPF~\cite{DBLP:journals/rfc/rfc2178}.  Intra-AS routing algorithms are chosen independently by each AS owner and generally designed to maximize efficiency and minimize network load. 

Inter-AS routing is performed via a single global system following the border gateway protocol (BGP)~\cite{DBLP:journals/rfc/rfc4271}. This protocol is used to find paths between two ASes, and is usually dominated by policy considerations (e.g., geographical, political, legal, economic, etc.) and real-world constraints.  For instance, ISPs typically avoid transiting traffic across others' ASes with whom they do not have concrete business arrangement in place.

\end{document}